\documentclass[preprint]{aastex631}

\shorttitle{The destiny of Dark Matter}
\shortauthors{Tracanna \& Hansen}
\graphicspath{{./}{figures/}}

\begin{document}

\title{The destiny of Dark Matter}

\author[0000-0002-0786-7307]{Fabiano Tracanna}
\affiliation{Dark Cosmology Centre, Niels Bohr Institute,
University of Copenhagen, \\
Jagtvej 155, Copenhagen 2100, Denmark}

\author[0000-0002-4691-3935]{Steen H. Hansen}
\affiliation{Dark Cosmology Centre, Niels Bohr Institute,
University of Copenhagen, \\
Jagtvej 155, Copenhagen 2100, Denmark}



\begin{abstract}
The majority of baryons, which account for $15\%$ of the matter in the
Universe, will end their lives as carbon and oxygen inside cold black
dwarfs.  Dark matter (DM) makes up the remaining $85\%$ of the matter
in the universe, however, the fate of DM is unknown.  Here we show
that the destiny of
purely gravitationally interacting 
DM particles follows one of two possible routes.  The first possible
route, the "radiation-destiny" scenario, is that massive DM particles
lose sufficient energy through gravitational radiation causing them to
spiral into a supermassive black hole that ultimately disappears
through Hawking radiation. The second possible route, the
"drifting-alone" destiny, applies to lighter DM particles, where only
the central DM halo region spirals into the central BH which is then
Hawking radiated away. The rest of the DM halo is ripped apart by the
accelerated expansion of the Universe.
\end{abstract}



\section{Introduction}
\label{sec1}

In approximately 5 billion years, our Sun will evolve into a red giant, expanding its radius by several hundred times and engulfing the innermost planets of the solar system, likely including the Earth, which will become a scorched and lifeless desert. During the same period, the Milky Way, our galaxy, will collide with the Andromeda galaxy, forming a resulting galaxy named Milkomeda. This new galaxy will continue as a large elliptical galaxy, with its central black holes merging into a supermassive BH~\citep{2020A&A...642A..30S}. Due to the accelerated expansion of the Universe, only a few more galaxies will collide with Milkomeda, and after  several tens of billions of years, all solar-mass stars will fade away, and after trillions of years, all low-mass stars will have exhausted their fuel~\citep{1997RvMP...69..337A}.

The Universe comprises not only stars and gas but also large amounts of dark matter (DM). 
This has been observed in galaxies and clusters from the early 1930th
\citep{1930MeLuF.125....1L, 1932BAN.....6..249O, 1937ApJ....86..217Z}, 
thoroughly established with galactic rotations curves in the 1970th
\citep{1980ApJ...238..471R},
and confirmed on the scales of the full universe through the
cosmic microwave background observations \citep{2016A&A...594A..13P}.

DM particles have a negligible collisional cross-section \citep{2004ApJ...606..819M}, which implies that they orbit the galaxy solely under the influence of gravity. 
The ultimate fate of DM particles depends on their annihilation rate or
whether they decay.
In some of the most popular particle scenarios, DM can annihilate when two DM particles come close enough to each other~\citep{2005PhR...405..279B} . These models establish a strong correlation between the DM annihilation cross-section and the DM abundance. A typical annihilation cross-section for such particles is of the order of $\sigma \sim 10^{-32}  
\left( \frac{m}{1{\rm TeV}} \right)^2 {\rm cm}^{2}$,
where the mass of $1 {\rm TeV}$ is often considered for thermally produced DM particles~\citep{2005PhR...405..279B}
This means that a significant fraction of DM in a typical galaxy with a mass of $10^{12} M_\odot$ will annihilate within $10^{17}$ years.

During DM annihilation, the resulting products may include photons and other high-speed particles that escape the galaxy. As more DM particles annihilate, the central part of the galaxy vanishes. This causes a reduction in the gravitational attraction, allowing DM particle velocities to exceed the escape velocity when $\sim 10\%$ of the DM particles have annihilated  \citep{2008gady.book.....B} (see Appendix A).
As a result, if DM particles are thermally produced, and hence typically
have annihilation cross sections of the order the weak interaction scale, 
then the DM in the galaxy disperse into the emptiness of the expanding space. The ultimate fate of these dispersed DM particles depends on their specific particle properties, including whether they decay into lighter particles or not. 
Other popular particle candidates for the DM 
considers decaying DM particle. For instance
for the sterile neutrino \citep{1994PhRvL..72...17D}
the decay time into active neutrinos or photons
is approximately $\tau_{\rm decay} \sim 10^{19} \, \left( 
\frac{10 {\rm keV}}{m}\right)^3$ years. As the particles
decay away the resulting less tightly bound galaxy will 
be ripped apart by the accelerated expansion of space, 
and the dispersed DM particles will decay.
We will not consider neither thermally created DM particles
or sterile neutrinos further.

In contrast, the destiny of the more general class of DM particles that only interact through gravity is generally unknown, and we will here show how it depends sensitively on their particle mass.

There is a long history of  production of particles 
with no non-gravitational interactions 
\citep{1969PhRv..183.1057P, Grib:1969ruc, Parker:1971pt, Mamaev:1976zb, Grib:1976pw}, and this production may even lead to abundances relevant for it being the DM \citep{Dolgov:1989us, Traschen:1990sw, Dolgov:1998wz, Kuzmin:1998kk, Bassett:1997az, Chung:1998ua, Chung:2001cb}.
The DM can be created for instance by allowing it to have a coupling to a scalar inflaton field $\phi$, such as
${\mathcal L} \sim {\phi}^2  {\chi}^2$, 
where the DM is a scalar $\chi$. The inflaton field oscillates towards the end of inflation, and the DM is produced due to the nonadiabatic expansion of spacetime during the transition to the matter or radiation dominated phase. An important restriction is that the DM must have properties to prevent subsequent thermalization, which is often achieved by considering very massive DM particles with no other coupling to standard model particles.
Alternatively, the coupling of DM can be directly to gravity (and nothing else), through a conformal coupling like
${\mathcal{L}} \sim \xi R \chi^2$, 
where $R$ is the Ricci scalar curvature. The fundamental cause of particle production 
is that the expanding Universe breaks time-translation symmetry, which leads to non-conservation of energy in the quantum 
particles~\citep{Ford:2021syk}.
For recent lists of references, see for instance
\citep{Ford:2021syk, Lebedev:2021xey, 2023JHEP...05..181K}.
This wide range of production mechanisms have one thing
in common, namely that the DM particles today will appear essentially sterile and non-interacting, except through
gravity.
Thus, for the rest of this paper, we are only considering the general
class of DM particles that today only interact through gravity.

Massive objects circling each other have been
predicted to emit gravitational waves (GW) \citep{Einstein:1918btx}. 
This effect was
measured indirectly by the frequency change of pulsars
\citep{1981SciAm.245d..74W}, and more recently 100 years of 
search culminated with the direct observation of GW
\citep{Abbott_2016}.

The purely gravitationally interacting DM particles emit gravitational waves as they orbit the galaxy, and the power emitted by a single particle can be expressed as
$P_m(r) = \frac{32G}{5c^5} \Omega^6 m^2 r^4$,
where $G$ is the gravitational constant, $c$ is the speed of light,
$\Omega$ is the angular velocity of the particle, and $r$ is
the radius of its orbit (see appendix B). 
Notably, the power emitted depends on the square of the particle's mass. Thus, if we consider a wide range of DM candidates with masses ranging from that of the axion particle ($m \sim 10^{-38} g$) to a 100 solar-mass DM candidate ($m \sim 10^{35} g$), there will be a difference of a factor of $10^{146}$ in the emitted power.
The angular velocity of the DM particle depends strongly on the potential existence of a supermassive central black hole, since the typical circular velocity can be expressed as $v_{\rm circ}(r) = \sqrt{GM(r)/r}$. Therefore, a large central object will allow the DM to emit more gravitational radiation and transit to a smaller orbit at a faster rate.
A central BH is not a permanent fixture, as it is subject to Hawking radiation due to quantum effects~\citep{1975CMaPh..43..199H}. The timescale for a black hole to radiate away due to this effect is on the order of $\tau_{\rm Hawking} \sim 10^{-19} \left(\frac{M_{\rm BH}}{g}\right)^3$ years. Thus, a BH with a mass of $10^{6}M_{\odot}$ will completely evaporate in approximately $10^{85}$ years.

\begin{figure}[ht]
    \centering
    \includegraphics[width=0.8\textwidth]{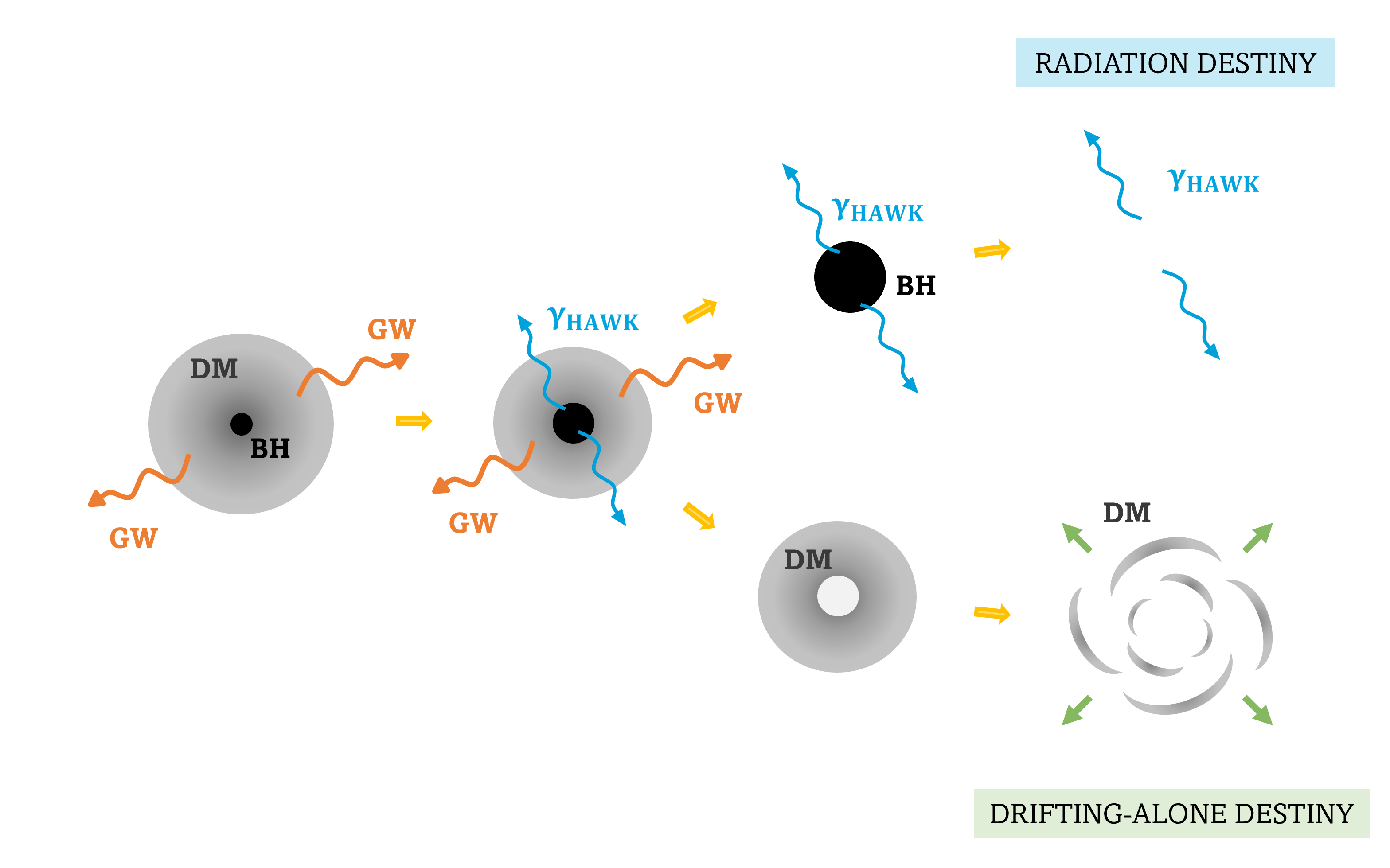}
    \caption{The two main evolutionary tracks of dark matter haloes: 
    for very massive DM candidates the large amount of 
    gravitational radiation emitted 
    leads to a quick inspiral onto the central
    BH. Subsequently this BH evaporates, and all DM thereby disappear in
    radiation. Alternatively, for light DM candidates, the central
    BH evaporates before a significant fraction of the DM has collapsed
    onto the central BH, and subsequently the remaining part of the
    DM halo will be dispersed into the expanding Universe.}
    \label{fig:big picture}
\end{figure}

\section{The two possible destinies}
\label{sec:results}
For a given DM particle mass, $m$, we can  calculate the timescale for a fraction of the galaxy to inspiral due to energy loss from gravitational radiation. For instance, we can ask how long it would take for the innermost $10^{-9}$ of the galaxy's mass to move on sufficiently small orbits that the DM particles will get absorbed by the central BH. For a $10^{12} M_\odot$ galaxy with a density profile in reasonable agreement with observations and numerical simulations \citep{Hernquist} (see Appendix B2), and with an initial central BH of $10^{6} M_\odot$, this timescale is approximately $10^{80}$ years for a DM particle with mass $m=100 {\rm TeV}$. This is a shorter time than the Hawking radiation timescale of $10^{85} {\rm yrs}$ discussed above.
For a DM candidate with mass $m = 10^{-5} {\rm eV}$, the corresponding timescale is on the order of $10^{118}$ years. Thus, the destiny of DM particles in a $10^{12} M_\odot$ galaxy with a central BH of $10^{6} M_\odot$ is fundamentally different for different DM particle masses. The most massive DM candidates will lose enough energy through gravitational radiation to be entirely absorbed by the central BH in a sufficiently short time, and their fate is to end as Hawking radiation on timescales of the order of $10^{103}$ years. This is shown as "radiation-destiny" in Figure 1. On the other hand, lighter DM candidates will lose energy through gravitational radiation slowly enough that the central BH will evaporate. Subsequently, the remaining galaxy will be dispersed into the vast empty space. In this case, shown as "drifting-alone destiny" in Figure 1, a significant fraction of the individual DM particles will survive.

\begin{figure}
    \centering
    \includegraphics[width=0.8\textwidth]{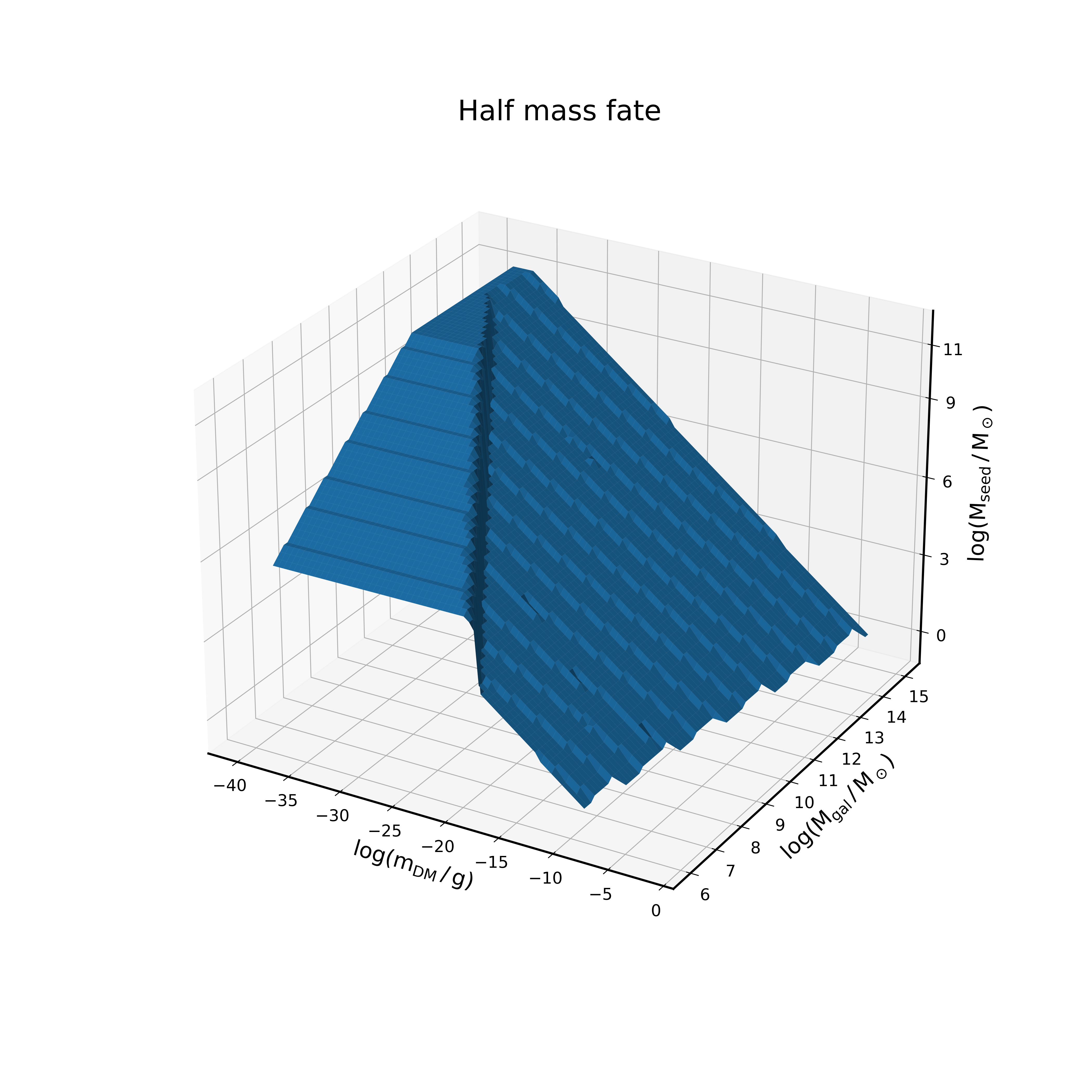}
    \caption{The figure illustrates the fate of dark matter for a wide range of galaxy parameters. The blue-colored
surface and the region under it represent the parameter space in which the central black hole evaporates rapidly enough that at least half of the dark matter particles in the cosmological structure end up dispersed in the Universe. Conversely, the non-colored region shows parameters for which more than half of the structure ends up being engulfed by the central black hole, which subsequently evaporates through Hawking radiation.
    The calculation spans over 70 orders of magnitude in the dark matter particle mass, $m_{\rm DM}$, and galaxy masses ranging from dwarf galaxies of $10^6 M_\odot$ to galaxy clusters of $10^{15} M_\odot$.
    It also allows for the possibility of the central seed black hole having a range of masses, from a single star to all the stars of the structure.
    The figure is cut short at  ${\rm log} 
    \left( m_{\rm DM}/g \right) \sim 0 $ to better visualize the surface, as heavier dark matter particles
 lie out and above the blue-colored surface for any galaxy and seed black hole mass shown.
    }
    \label{fig:overviewfate}
\end{figure}

\section{Three parameters $M_{\rm gal}$, $M_{\rm BH}$ and $m_{\rm DM}$}\label{sec12}

As DM particles are treated as point-like, they only interact through 2-body gravitational interactions that are long-range. The corresponding relaxation process can affect their energy distribution, which may result in the ejection of DM particles if their velocities exceed the local escape velocity \citep{1940MNRAS.100..396S}. However, there is a counteracting effect of dynamical friction \citep{1943ApJ....97..255C}, which reduces the velocity of the fastest particles. Since relaxation is a stochastic process and dynamical friction provides a systematic deceleration, the resulting energy distribution of DM particles will not contain particles that will evaporate from the cosmological structure (see appendix C for more details).

Galaxies can be distinguished based on the available gas and stellar matter, the total mass of the dark matter halo, and the mass of individual dark matter particles. Additionally, the initial mass of the central object can vary from a single stellar mass to the mass of the Milky Way's black hole, which is approximately $10^6 M_\odot$, or to supermassive black holes with masses exceeding several $10^9 M_\odot$.
We have today observed a free-floating BH of mass $\sim 7 M_\odot$
\citep{Sahu_2022} and EHT took the first image of
the black hole at the center of galaxy Messier 87
\citep{2019ApJ...875L...1E}.
The Milky Way has a BH of mass
$4 \cdot 10^6 M_\odot$ \citep{1998ApJ...509..678G, Schodel:2002py}, 
Andromeda has a BH of mass $\sim 1.4 \cdot 10^8 M_\odot$ \citep{10.1063/5.0027838}, and it is believed that 
most massive galaxies host a supermassive BH 
near its center \citep{2013ARA&A..51..511K}.

The gas and stellar matter will either be ejected from the galaxy or absorbed by the central black hole on much shorter timescales than those of the dark matter, and to 
avoid the details of this complication here, we simply allow these
options to be covered by the seed BH mass to range from
stellar mass to the entire mass of the cosmological 
structure (see Appendix C).~\footnote{In
principle, one should also consider collisions between
stars and the DM particles, in the case of sufficiently massive
DM particles. For DM particles not thermally created one can
still assume to ignore the non-gravitational scattering and
hence just include gravitational interactions.
In this case the physical extend of the DM particle 
becomes relevant, since a point-like DM particle with solar-mass
could exhibit strong energy exchange with the white (or cold, black) dwarfs. It is worth remembering that DM substructures 
created from much smaller individual DM particles are very large, 
typically of the order $10^{-2}$ pc for Earth mass DM haloes, and
about $10$ pc for solar-mass DM haloes
\citep{Diemand:2009bm, Wang_2020}. 
A few of these substructures typically survive the formation of the
galaxies, and will hence also affect the stellar motion.
For simplicity, we will 
ignore the effects of interactions between the black dwarfs and the
DM particles in this paper.}
Consequently, we can reduce the number of important parameters to three: the mass of the seed black hole (which may include all the mass of present-day stars and gas), the total mass of the cosmological structure, and the mass of the 
individual dark matter particle.

In Figure \ref{fig:overviewfate}, we present the fate calculation for a wide range of possible parameters. The general conclusion is that dark matter survives, i.e., it does not get absorbed by the central black hole, for small dark matter particle masses. At the second order, smaller seed black hole masses allow for more dark matter to survive.

\section{Conclusion}
The properties of DM particles are mostly unknown, and they may potentially decay or undergo annihilation. This paper examines the scenario where DM particles solely interact gravitationally over significantly longer timescales compared to the current age of the Universe.

We have demonstrated that the fate of DM is highly dependent on the mass of the DM particles. Extremely massive DM particles will promptly emit gravitational waves, leading to their gradual spiral towards the central black hole of the galaxies. Subsequently, the black hole will emit Hawking radiation, causing the DM particles to ultimately disappear as radiation.

In the case of lighter DM particles, the emission of gravitational radiation occurs at a significantly slower rate. As a result, only a minor portion of the DM particles will be absorbed by the central black hole. Once the central black hole ceases to exist, the potential of the galaxy is slightly reduced, and 
the remaining DM particles within these cosmological structures will gradually evaporate. Consequently, these DM
particles will follow the "drifting-alone" destiny.









\color{black}

\appendix

\section{Evaporation of DM particles when central region
has annihilated away}

\color{black}
As discussed in the introduction, some DM particle candidates have annihilation cross section of the order the weak interaction scale.
Such particle most often also annihilate when two particles get
close to each other.
\color{black}

When DM particles have a non-zero annihilation
cross section, then the central part of the halo will 
first disappear since the annihilation rate is proportional
to $\rho^2$. In this case the potential of the structure is
reduced, and hence the high-energy tail of the DM distribution
function may evaporate from the halo. The corresponding
calculation is as follows.

\begin{figure}[ht]
    \centering
    \includegraphics[width=0.8\textwidth]{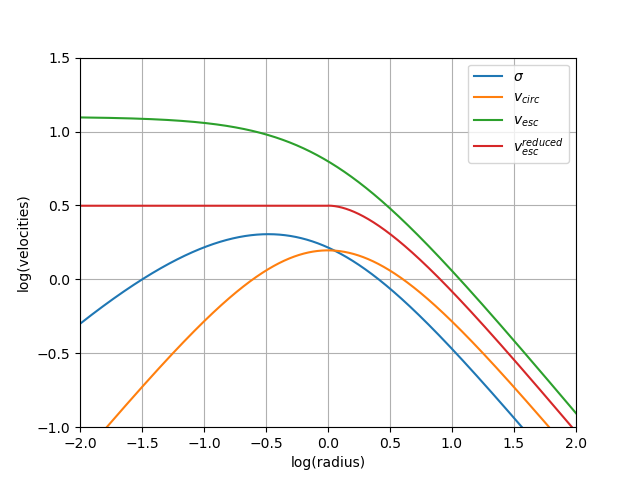}
    \caption{Velocities as a function of radius in a typical DM halo,
    with scaled distances and velocities.
    The total velocity dispersion is given by $\sigma$ (blue line)
    the circular velocity is $v_{circ}$ (orange line). The uppermost
    curve is the escape velocity (green line) for the full structure, and
    the reduced escape velocity  (second from top, red line) is
    calculated when removing the central 
    total mass. When $v_{esc}/\sigma \approx 2$ at  a given radius, then
    the highest energy particles will escape, 
    and through a run-away process the
    entire structure will disperse.}
    \label{fig:vel}
\end{figure}

Consider a halo in equilibrium, with a density profile given by
$\rho (r)$. One can integrate the Jeans
equation to show that the radial velocity velocity dispersion
is given by
\begin{equation}
    \sigma_r^2 (r) = \frac{1}{\rho(r)} \int_r ^\infty
    \frac{\rho (r') G M(r')}{r'^2} \, dr' \, 
\end{equation}
From numerical simulations it is known that the velocity distribution
function does not have an exponential tail, but instead has
a rapid decline which goes to zero around $v = 2 \sigma_{\rm tot}$
\citep{2006JCAP...01..014H}. Thus, if the escape velocity
\begin{equation}
    v_{\rm esc} (r) = \sqrt{-2\Phi (r)} \,
\end{equation}
where $\Phi(r)$ is the potential of the structure, becomes
smaller than approximately 2 times the total velocity dispersion,
then the high-energy particles will escape. Assuming that the
velocity anisotropy is zero, one has $\sigma_{\rm tot}^2 = 3 \sigma_r^2$,
and we find that if $10\%$ of the central mass (in a Hernquist 
structure) is removed, then the potential is reduced by slightly more
than $10\%$ at all radii. This will reduce the potential even further,
leading to a run-away process where all the DM particles will evaporate. This is shown in figure \ref{fig:vel}.

Another popular particle candidate for DM is a decaying DM particle. For instance, in the case of sterile neutrinos \citep{1994PhRvL..72...17D}, the decay time into active neutrinos or photons is approximately $\tau_{\rm decay} \sim 10^{19} \, \left( \frac{10 {\rm keV}}{m}\right)^3$ years. As these particles decay, the resulting galaxy becomes less tightly bound, 
some of the particles now have velocities exceeding the escape velocity
of the galaxy, and eventually the remaining part of the galaxy is ripped apart by the accelerated expansion of space. The dispersed DM particles will also decay.

\section{Linearized gravity}
\label{sec:Lin_grav_details}
To describe the GW emission we work under the assumptions of linearized gravity. This amounts to considering small GW amplitudes, large distances from the source and short wavelength GWs. 

Linearized gravity implies assuming that gravitational waves (GWs) are a small perturbation to the Minkowski metric $\eta_{\alpha \beta} \equiv \text{diag}(-1, 1, 1, 1)$.
The GWs will therefore be described by a metric perturbation $h_{\alpha \beta}$ such that the metric solving the Einstein equation can be written as 
\begin{equation}
    g_{\alpha \beta} = \eta_{\alpha \beta} + h_{\alpha \beta},
\end{equation}
with $h_{\alpha \beta} \ll 1$ for every $\alpha, \beta$.

In linearized gravity, the Einstein equation assumes the form
\begin{equation}
    \square \bar{h}_{\alpha \beta} = -16\pi T_{\alpha \beta}
    \label{eq:Einstein}
\end{equation}
\citep{Hartle},
in geometrized units ($ c = G = 1$, mass measured in length).
$T_{\alpha \beta}$ is the stress-energy tensor and the \textit{trace-reversed} amplitude $\bar{h}_{\alpha \beta}$ is defined by
\begin{equation}
    \bar{h}_{\alpha \beta} \equiv h_{\alpha \beta} -\frac{1}{2}\eta_{\alpha \beta} h \,,
    \label{eq:h_bar}
\end{equation}
with $h$ being the trace of the metric perturbation (i.e. $ h \equiv h^{\gamma}_{\gamma}$), and $\eta_{\alpha \beta}$ being the Minkowski metric.
The d'Alembert operator $\square$ is defined as $$\square \equiv \frac{\partial}{\partial x_{\nu}} \frac{\partial}{\partial x^{\nu}} = -\frac{\partial^2}{\partial t^2} + \nabla^2 ,$$
following the convention ($-$ + + +) for the metric signature.  
Imposing gauge conditions allow to close the system and uniquely solve the equation. We choose the Lorenz gauge, that can be conveniently expressed in terms of $\bar{h}_{\alpha \beta}$ as 
\begin{equation}
    \frac{\partial \bar{h}^{\alpha \beta}}{\partial x^{\beta}} = 0 \, .
    \label{eq:lorenz}
\end{equation}
It can be shown \citep{Hartle} that the spatial components of the trace-reversed GW amplitude can be written as 
\begin{equation}
    \bar{h}^{ij}(t, \vec{x}) \longrightarrow \frac{2}{r} \ddot I^{ij}(t-r),
    \label{eq:amplitude_linearized}
\end{equation}
where the second mass moment $I^{ij}$, here evaluated at the retarded time $t - r/c = t- r$, is defined as
\begin{equation}
    I^{ij} \equiv \int x'^i x'^j \rho(t, \vec{x}) d^3x'.
    \label{eq:second_mass_moment}
\end{equation}
The energy flux (energy per unit time per unit area) $f_{GW}$ of a linearized, plane GW is proportional to the square of the amplitude of the GW, let us call it $a$, times the square of its frequency $\omega$ \citep{Hartle}:
\begin{equation}
    f_{GW} = \frac{\omega^2 a^2}{32\pi}.
    \label{eq:energy_density_GW}
\end{equation}
Since we are looking at the GW far away from the source, and the amplitude in Eq.~\ref{eq:amplitude_linearized} describes a spherical wave, the plane wave approximation is legitimate. The frequency dependence in Eq. \ref{eq:energy_density_GW}, together with Eq. \ref{eq:amplitude_linearized}, suggests a dependence of $f_{GW}$ which is quadratic in the third time derivative of $I^{ij}$. We can write:
\begin{equation}
    f_{GW} \propto \frac{1}{r^2} \Big[\xi\Big(\dddot I^{ij}\Big)\Big]^2.
\end{equation}
The right function $\xi$ can be found by noticing that there is no radiation from a spherically symmetric mass distribution. The quadrupole moment tensor
can be expressed in terms of the second mass moment as
\begin{equation}
    Q^{ij} = 3I^{ij} - \delta^{ij} I^k_k,   
    \label{eq:quadrupole_mass_moment}
\end{equation}
satisfies such requirements. The total power radiated can be found by integrating $f_{GW}$ over a surface encompassing the mass distribution, say a sphere, in the limit $r\longrightarrow \infty$, i.e.
\begin{equation}
    P_{GW} = \lim_{r \to \infty} 4\pi \int f_{GW}r^2 dr \propto \dddot Q_{ij} \dddot Q^{ij}
\end{equation}
Including units  we can finally express the total power radiated by gravitational waves in the quadrupole approximation as
\begin{equation}
    P_{GW} = \frac{G}{45 c^5} \left< \dddot Q_{ij} \dddot Q^{ij} \right>,
    \label{eq:quadrupole_radiation}
\end{equation}
with $\left< {} \cdot {} \right>$ denoting a time average over a period \citep{Hartle}.

\subsection{Test mass in central gravitational field}
\label{sec:test_mass}
A DM mass $m$ is orbiting a central mass $M$, with $M \gg m$, in an elliptical orbit with such a low eccentricity that we can assume the orbit to be circular. Let $R$ be the initial radius of the orbit, $\Omega$ the orbital frequency. By placing the origin of our Cartesian coordinate system to coincide with the position of the central object and choosing the orbit to lie the $xy$ plane we can describe the trajectory of the test mass as:


\begin{eqnarray}
        x(t) &=& R \cos(\Omega t)\\
        y(t) &=& R \sin(\Omega t) \\
        z(t) &=& 0
\end{eqnarray}



The mass density of the system can be written as
\begin{equation}
    \rho(\vec{x}) = M\delta(\vec{x}) + m\delta(\vec{x} - \vec{r}),
\end{equation}
and the components of the second mass moment can then be written
\begin{eqnarray}
        I^{xx} &=& mR^2 \cos^2(\Omega t) = \frac{1}{2} mR^2 [1 + \cos(2\Omega t)]\\
        I^{xy} &=& mR^2 \cos(\Omega t)\sin(\Omega t) = \frac{1}{2} mR^2 \sin(2\Omega t)\\
        I^{yy} &=& mR^2 \sin^2(\Omega t) = \frac{1}{2} mR^2 [1 - \cos(2\Omega t)] \\
        I^{xz} &=& I^{yz} = I^{zz} = 0
        .
    \label{eq:test_mass_I_ij}
\end{eqnarray}
The remaining components are determined by the fact that the second mass moment is by definition a symmetric tensor, i.e., $I^{ij} = I^{ji}$.
The third time derivative of each non-zero component is easily calculated to be
\begin{eqnarray}
       \dddot I^{xx} &=& -4\Omega^3mR^2 \sin(2\Omega t) \\
       \dddot I^{xy} &=& 4\Omega^3mR^2 \cos(2\Omega t)\\
        \dddot I^{yy} &=& 4\Omega^3mR^2 \sin(2\Omega t) = -\dddot I^{xx} 
        .
    \label{eq:test_mass_I_ij_dddot}
\end{eqnarray}
Recalling the definition of the quadrupole moment \ref{eq:quadrupole_mass_moment} and noticing $\dddot{I} ^{kk} = 0$, since $I^k_k=mR^2$ is independent of time,
\begin{equation}
    \dddot Q_{ij} \dddot Q^{ij} = 144\Omega^6m^2R^4\big[\sin^2(2\Omega t) + 2\cos^2(2\Omega t) + \sin^2(2\Omega t) \big] = 288 \Omega^6 m^2 R^4.
\end{equation}
The power radiated \ref{eq:quadrupole_radiation} is thus
\begin{equation}
    P_{GW} = \frac{32G}{5c^5} \Omega^6 m^2 R^4.
    \label{eq:test_mass_power}
\end{equation}
This is in agreement with the results of \citep{1972gcpa.book.....W}.

\subsection{The mass profile of galaxies}
We assume the DM mass distribution to be spherically symmetric, 
with a density $\rho(r)$ described by the Hernquist profile \citep{Hernquist}
\begin{equation}
    \rho(r) = \frac{M}{2\pi} \frac{a}{r} \frac{1}{(r+a)^3},
    \label{eq:hernquist_profile}
\end{equation}
where $M$ is the total mass and $a$ a linear scale of the object. 
This profile approximates well the mass distribution of galactic bulges and elliptical galaxies, but also the DM distribution in haloes. We opt for this profile instead of the NFW profile
since the Hernquist mass is finite without the need
for a truncation at large radii. 
The cumulative mass profile, and the corresponding potential pertaining to the density profile \ref{eq:hernquist_profile} are, respectively \citep{Hernquist},
\begin{eqnarray}
        M(r) &=& M \frac{r^2}{(r+a)^2}, \\
    \phi(r) &=& -\frac{GM}{r+a}.
    \label{eq:hernquist_potential}
\end{eqnarray}


The velocity dispersion $\sigma_v^2$ is obtained by solving the 1D Jeans equation for a non-rotating, spherical system. Its radial component, $\sigma_{v_r}^2$, is given by \citep{Hernquist}
\begin{eqnarray}
            \sigma_{v_r}^2 &=&  \frac{GM}{12a} \Biggl\{ \frac{12r(r+a)^3}{a^4} \ln{\left(\frac{r+a}{r}\right)} \\
            &&- \frac{r}{r+a}\Biggl[ 25 + 52 \frac{r}{a} + 42 \biggl( \frac{r}{a} \biggr)^2 + 12\biggl( \frac{r}{a} \biggr)^3 \Biggr] \Biggr\}.
    \label{eq:sigma2}
\end{eqnarray}
The corresponding angular velocity is given by 
\begin{equation}
    \Omega^2 = \frac{\sigma_{v_T}^2}{r^2} = \frac{2\sigma_{v_r}^2}{r^2},
\end{equation}
with $\sigma_{v_T}^2 \equiv \sigma_{v_\theta}^2 + \sigma_{v_\phi}^2$ being the transverse velocity dispersion. In fact, for a spherical system, $\sigma_{v_r}^2 = \sigma_{v_\theta}^2 = \sigma_{v_\phi}^2$, so that $\sigma_{v_T}^2 = 2 \sigma_{v_r}^2$. The single particle GW power loss is thus given by
\begin{equation}
    P_{\rm m}(r) = \frac{32Gm^2}{5c^5} \frac{8 \bigl(\sigma_{v_r}^2 \bigr)^3}{r^2}.
    \label{eq:single_particle_power_sigma_T}
\end{equation}



\begin{figure}
    \centering
    \includegraphics[width=0.8\textwidth]{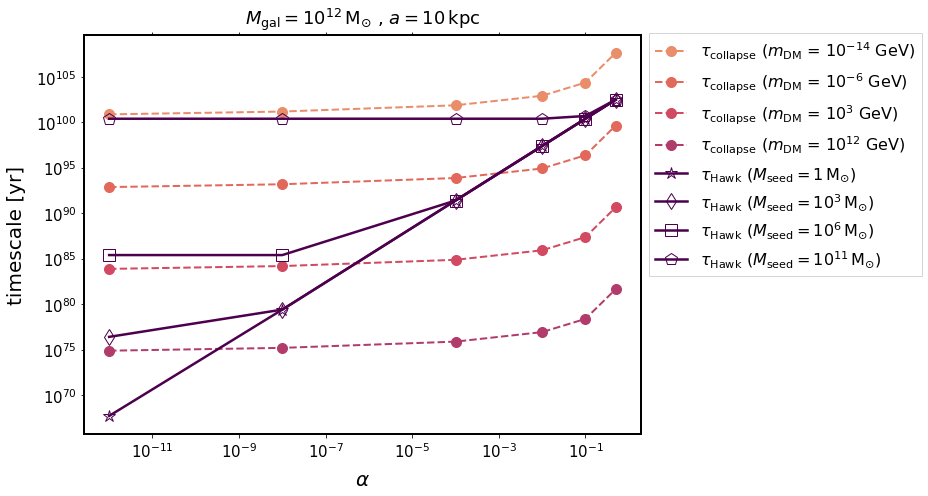}
    \caption{Timescales for inspiral and Hawking radiation as a function
    of the fraction of the galaxy. The four dashed lines show the dependence on
    the mass of the DM particle, where the upper-most curve are for the
    lightest DM particle ($m_{\rm DM} = 10^{-14} {\rm GeV})$ and the lowest
    curve is the most massive ($m_{\rm DM} = 10^{12} {\rm GeV})$. The four
    solid lines show the Hawking radiation timescale dependence of the 
    initial seed BH mass. If the initial seed BH is small (lowest curve,
    $1 M_\odot$)
    then the Hawking radiation timescale is short, whereas a supermassive
    BH initial seed of $10^{11} M_\odot$ leads to very long radiation
    timescales (uppermost solid curve).}
    \label{fig:timescales}
\end{figure}

In figure \ref{fig:timescales} we show
that for a DM particle of mass $1TeV$ and an initial seed BH
of mass $10^6 M_\odot$ the Hawking radiation timescale is longer
than the inspiral time for the entire galaxy, and hence this 
structure will inspiral and gets absorbed by the central BH. 
In contrast
a central initial seed BH of only $10^3 M_\odot$ will evaporate
away before a fraction $10^{-6}$ of the galaxy has inspiraled
onto the BH. This implies that all subsequent inspiraling DM will 
eventually evaporate through Hawking radiation, until the
remaining DM halo is sufficiently dilute that it will disperse
through the accelerated expansion of the Universe, and hence
a significant fraction of the DM particles will remain 
as DM particles in the
expanding Universe.

\newpage

\section{Evaporation v.s. Dynamical Friction}\label{secA1}

It has almost become "common knowledge" that gravitational 2-body interactions 
lead to effectively relaxed systems, which implies a slow but steady
evaporation of particles from the system \citep{1940MNRAS.100..396S}. 
The argument is that the 2-body
interactions leads to an exponential distribution of energies, and since
any exponential will have a high-energy tail beyond the systems escape-velocity,
then this implies that particles will evaporate. This conclusion is, however, 
incorrect, as we will show now, since it ignores another important gravitational
effect: Dynamical Friction (DF) \citep{1943ApJ....97..255C}.

The relaxation time 
arises from long-range encounters causing a cumulative 
diffusion of a stars velocity.
It is frequently estimated by following the trajectory of
a subject star with initial velocity $v$, as it passes a field star with
impact parameter $b$. 
The acceleration from the field star gives the subject star a perpendicular
velocity of the order $\delta v = 2Gm/(bv)$ \citep{2008gady.book.....B}. 
If we consider a large
spherical structure with radius $R$ and $N$ particles each with mass $m$,
then we can calculate the number of long-range encounters during one
crossing. Each encounter produces a small perturbation to the subject stars velocity, 
and since these
are independent of each other we can add the $\delta v^2$ linearly. Hereby
one can integrate over all impact parameters to find
\begin{equation}
    \Delta v^2 \approx 8 N \left( \frac{Gm}{Rv}\right) ^2 \, {\rm log} \Lambda \,
\end{equation}
where the Coulomb logarithm comes from the maximum and minimum impact
parameters $b_{\rm max} \sim R$ and $b_{\rm min} \sim R/N$, giving ${\rm log} \Lambda \sim {\rm log} N$. 
It is important to keep in mind that the standard trick of numerical N-body
simulations of the inclusion of a softening merely leads to a slightly bigger
value for $b_{\rm min} $, which only enters the expression through the
${\rm log} \Lambda$.
A typical velocity is given by
\begin{equation}
    v^2 = \frac{GNm}{R} \, ,
\end{equation}
and we hence have
\begin{equation}
    \frac{\Delta v^2}{v^2} \approx \frac{8 \, {\rm log} N}{N} \, ,
\label{eq:nrelax}
\end{equation}
which implies that after $\frac{N}{8 \, {\rm log} N}$ crossings
the totalt energy exchange is of the same level as the initial energy
(the stars orbit has been completely randomized),
and this gives the result
\begin{equation}
    t_{\rm relax} = \frac{N}{8 {\rm log} N} \, t_{\rm cross} \, . 
\end{equation}
This effect is possibly most famous for Globular clusters, where $N\sim 10^5$
and crossing times of Myrs makes this 2-body relaxation important given the
age of the globular clusters.
If these repeated encounters set up a Maxwellian distribution of velocities, 
then the high-energy tail will contain particles moving beyond the
esacape velocity, and these particles will hence evaporate. Given
the small number of particles in the high-energy tail, one often expects that the
entire cosmological structure may evaporate at
time-scales around 100 times the relaxation time \citep{1940MNRAS.100..396S}.

There is, however, another gravitational effect, which also must be 
included, namely the Dynamical Friction (DF). This effect is often interpreted
through the gravitational focusing behind the particles path, which slows
the particle down, and hence transfers energy from the rapidly moving particles
to the slow ones. By integrating over impact parameters the acceleration is
often written by Chandrasekhars expression \citep{1943ApJ....97..255C}
\begin{equation}
    \frac{d \vec v_M}{dt} = - 16 \pi^2 \, G^2 m \, \left(m+M \right) \,
    {\rm log} N \, \frac{\vec v_M}{v_M^3} \, \int _0 ^{v_M} f(v_m)v_m^2 dv_m \, ,
\end{equation}
where the subject star has mass $M$ and the field stars have mass $m$. From
this formula
it is clear that only the slower moving field particle contribute to slowing
the subject particle down. For a rapidly moving subject particle the integral 
over the field particles is just the number density 
$\int _0 ^{\infty} f(v_m)v_m^2 dv_m = n/(4\pi)$\, and hence the magnitude of the
acceleration can be written as 
\begin{equation}
    \frac{dv_M}{dt} = 8 \pi \left( \frac{Gm}{v_M}\right)^2 \,  {\rm log}N \, n
\end{equation}
where we used $M=m$ when considering only DM particles. 
To make the comparison with the relaxation time as explicit as possible
we will again consider a sphere 
of radius $R$ with $N$ particles of mass $m$, where a typical velocity is still
given
by $v^2 = GmN/R$. If we are considering a fast moving particle, then 
we can ask the number of crossings (of crossing time $\tau _{\rm cross} = R/N$) the
particle needs, in order to reduce its velocity by the order $v$
\begin{equation}
    \frac{dv_M}{dt} \, n_{\rm cross} \, \tau _{\rm cross} \approx v \, ,
\end{equation}
which is solved by
\begin{equation}
    n_{\rm cross}^{-1} \approx \frac{6 \, {\rm log} N}{N}.
\end{equation}
Comparing with eq.~{\ref{eq:nrelax}}
we thus see that the timescale for reducing the velocity of fastmoving
particles is the same (within a factor of 3/4) as that of evaporation.

The process of relaxation/evaporation is a stochastic process, whereas
DF has a systematic decelerating effect. Any given particle which happens
to have a velocity slightly larger than the field particles will 
therefore have its velocity reduced by DF faster than the statistical
process of relaxation can push it beyond the escape velocity.

The inclusion of DF in the calculation of stellar evaporation was first
studied in 
\citep{1943ApJ....97..263C} by considering the stochastic process of
relaxation as a diffusion process. The conclusions of \citep{1943ApJ....97..263C}
was also that the effect of DF is crucial to include in order to calculate
evaporation, even though the paper \citep{1943ApJ....97..263C} works under
the assumption of Gaussian distributions of velocities, which is today
known to be incorrect long before the onset of effects of both relaxation of DF
\citep{2006JCAP...01..014H}. It is expected that the very rapid process of
violent relaxation \citep{2008gady.book.....B} is responsible for the 
appearance of the non-exponential shape of the velocity distribution
function with no high-energy particles.
As shown in Figure 4, the stochastic appearance of 
high-energy particles will immediately be
damped by DF, and hence no DM particles will evaporate.
This calculation only considers $N=10^3$ particles, and the effect of DF
is only calculated accurately for the high-energy tail of the 
energy distribution (the bulk of the particles have their energies
adjusted accordingly to assure energy conservation in each time-step),
and thus a more careful calculation is needed in order to address 
complicated dynamical systems like Globular clusters. The above
argument (and simple calculation) is here partly used as an argument
why we may allow the "initial seed" BH to cover everything from a
single star, to the mass of the entire collection of stars.

\begin{figure}[ht]
    \centering
    \includegraphics[width=1\textwidth]{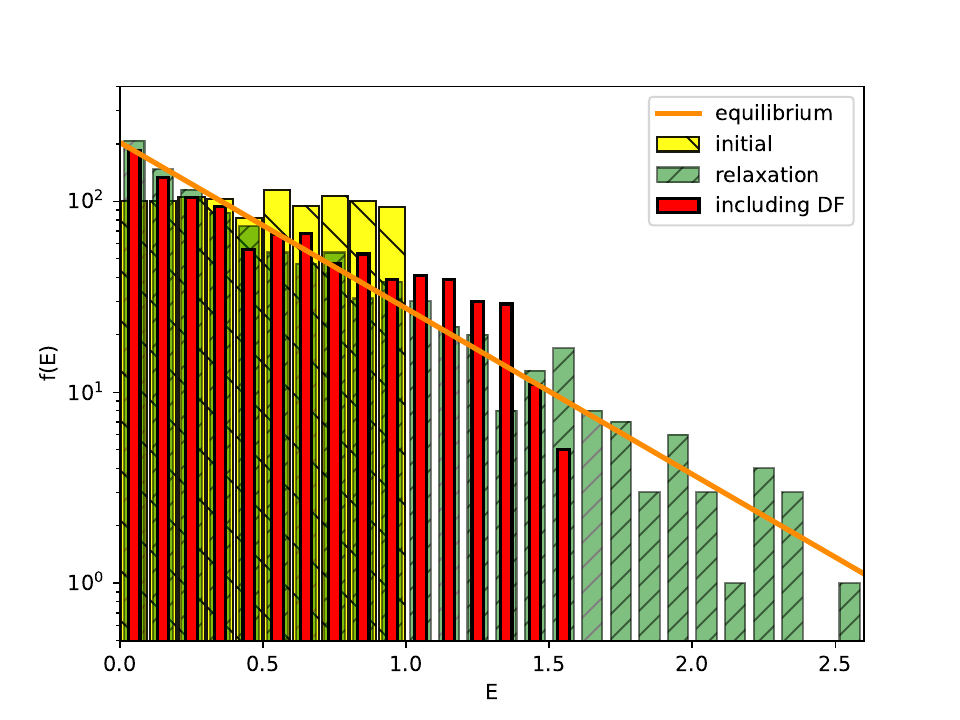}
    \caption{The normalized energy distribution. 
    The yellow dashed histogram
    shows an energy distribution of $N=10^3$ particles
    with random energies between 0 and 1. By allowing the
    particles to exchange energy through elastic collision
    one obtains the orange line analytically. Numerically
    by allowing
    $10^4$ times $N/2$ collisions with a maximal energy exchange of 
    $1\%$ of the particle energies, one gets the green 
    (up-sloping hatched) histogram, which follows the analytical
    prediction. By including the effect of dynamical friction
    for the $5\%$ of the particles with highest energy 
    (corresponding to the $2\sigma$ tail), one
    get a reduction of the high-energy tail, as represented by
    the solid red histogram. Here we included the factor of
    $3/4$ as derived above. Interestingly, it is visible how
    the high-energy particles are now slightly piled up
    at lower energies.}
    \label{fig:edf}
\end{figure}

\color{black}
\begin{acknowledgments}
It is a pleasure 
thanking the referee for very constructive suggestions which improved
the paper.  SHH thanks Jens Hjorth and Radek Wojtak
for interesting discussions.
\end{acknowledgments}
\color{black}

%






\bibliography{destinyaas3}{}
\bibliographystyle{aasjournal}



\end{document}